\documentclass[aps,prl,superscriptaddress,twocolumn,showpacs]{revtex4}
\usepackage{amsmath,bm}
\usepackage{graphicx,epsfig,braket,amssymb,amsfonts,color}

\newcommand{\ba}{\begin{eqnarray}}
\newcommand{\ea}{\end{eqnarray}}
\newcommand{\be}{\begin{equation}}
\newcommand{\ee}{\end{equation}}
\newcommand{\bd}{\begin{displaymath}}
\newcommand{\ed}{\end{displaymath}}
\renewcommand{\v}[1]{{\bf #1}}
\newcommand{\bpm}{\begin{pmatrix}}
\newcommand{\epm}{\end{pmatrix}}
\newcommand{\nn}{\nonumber \\}

\begin{document}

\title{Soliton Defects in One-dimensional Topological Three-band Hamiltonian}

\author{Gyungchoon Go}
\affiliation{Department of Physics, Sungkyunkwan University, Suwon
440-746, Korea}
\author{Kyeong Tae Kang}
\affiliation{Department of Physics, Sungkyunkwan University, Suwon
440-746, Korea}
\author{Jung Hoon Han}
\email[Electronic address:$~~$]{hanjh@skku.edu}
\affiliation{Department of Physics, Sungkyunkwan University, Suwon
440-746, Korea} \affiliation{Asia Pacific Center for Theoretical
Physics, POSTECH, Pohang, Gyeongbuk 790-784, Korea}

\begin{abstract} Defect formation in the one-dimensional topological three-band
model is examined within both lattice and continuum models. Classic results of
Jackiw-Rebbi and Rice-Mele for the soliton charge is generalized to the three-band model.
The presence of the central flat band in the three-band model makes the soliton charge as
a function of energy behave in a qualitatively different way from the two-band Dirac
model case. Quantum field-theoretical calculation of Goldstone and Wilczek is also
generalized to the three-band model to obtain the soliton charge. Diamond-chain lattice
is shown to be an ideal structure to host a topological three-band structure.
\end{abstract}
\maketitle

\begin{figure*}[ht]
\includegraphics[width=180mm]{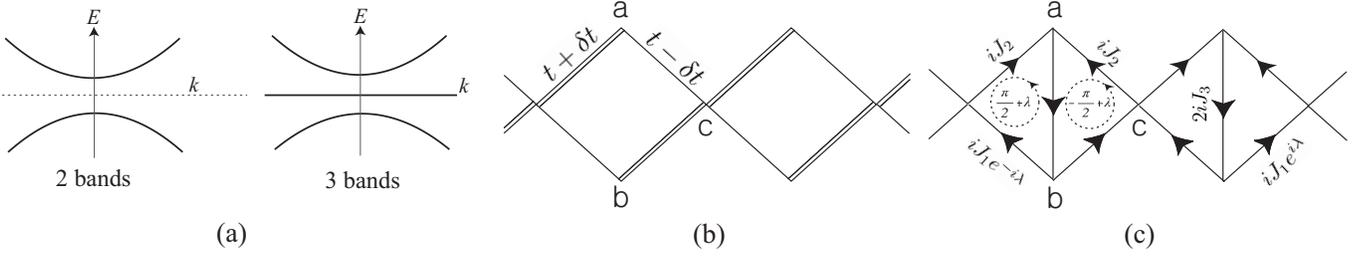}
\caption{(a) Schematic picture of the
generic two-band (left) vs. three-band (right) anti-crossing in a band structure. (b) The
ground state hopping configuration in the diamond-chain lattice model of $H^{\rm I}$. (c)
Hopping patterns for $H^{\rm II}$. Flux penetrating the two triangles are $\pm \pi/2
+\lambda$, respectively.} \label{fig:1}
\end{figure*}

\textit{Introduction}.- Defects are a useful way of un-earthing topological properties of the
underlying band structure~\cite{SSH,dhlee}. A classic example is the Su-Schrieffer-Heeger solitonic
defect formed at the kink of a Peierls insulator~\cite{SSH}. Its existence as the mid-gap state is
a manifestation of the massive Dirac nature of bulk one-dimensional band. Soliton nucleation due to
$\pi$-flux threading a two-dimensional topological band insulator is another example of the defect
revealing the topological nature of the bulk band~\cite{dhlee}. In all known instances so far,
however, the physics near the gap-opening momentum is treated within the two-component Dirac theory
drawn from the upper and the lower bands (Fig. \ref{fig:1}(a)). The well-known Jackiw-Rebbi
mechanism~\cite{Jackiw:1975fn} then automatically produces a soliton at the mass sign-changing
point. Later Jackiw-Rebbi model was generalized to treat cases without the particle-hole symmetry,
which gives rise to solitons carrying parameter-dependent irrational charge
\cite{RM,Kivelson,Jackiw:1982sm}. Generalization to cases involving more than two bands, however,
had not been made.

One may as well envision situations where the two bands are
``intervened" by a third one passing through the gapped region,
forming a three-band anti-crossing as schematically shown in Fig.
\ref{fig:1}(a). A minimal model, although not the most general
one~\cite{comment}, for such situation is the three-band Hamiltonian

\ba {\cal H}_{\v k} = \v d_{\v k} \cdot \v S \label{eq:d-dot-S-type}\ea
where $\v S = (S^x , S^y , S^z )$ are the three components of a spin-1 matrix and $\v d_{\v k}$ is
some momentum $(\v k)$-dependent vector. Its energy spectrum has one flat band, $\varepsilon_{\v k}
= 0$, in addition to a pair of symmetrically placed bands at $\varepsilon_{\v k} = \pm |\v d_{\v
k}|$~\cite{GPH,varma}. Such lattice model in two dimensions can be realized as complex-valued
tight-binding Kagome model at special fluxes~\cite{GPH}, or in the CuO$_2$ plane of the
cuprates~\cite{varma}. Bulk topological properties of two-dimensional lattice models of this sort
were examined in the past~\cite{OMN,varma,GPH}. In this paper, we address the nature of those
defects formed at the domain boundary in \textit{one-dimensional topological three-band models}.
\\

\textit{One-dimensional topological three-band model}.- We begin by addressing the question: what
kind of one-dimensional lattice Hamiltonians would map onto ${\cal H}_k = \v d_k \cdot \v S$ in
momentum ($k$) space? After experimenting with various possible structures we arrive at the
so-called diamond-chain lattice~\cite{vollhardt} as the most likely candidate supporting such
bands. Two routes can be followed to construct the model. One is by generalizing the Rice-Mele
two-band model~\cite{RM}. Schematics of such a lattice is depicted in Fig. \ref{fig:1}(b). Hopping
amplitudes along the sides are modulated as $t+\delta t$ (two lines) and $t-\delta t$ (one line).
On-site energies at the two opposing sites, labeled $a$ and $b$ in Fig. \ref{fig:1}(b), are
introduced as $\pm m$. There are four degenerate ground state configurations - two of which are
shown in Fig. \ref{fig:1}(b) - in such a model. Choosing the vacuum configuration shown on the left
of Fig. \ref{fig:1}(b), for instance, yields the $k$-space Hamiltonian

\begin{align}\label{DCham}
H^\mathrm{I}&=\sum_k \Psi_k^\dag \left[2\Bigl(\sqrt 2\, t\cos k ,\sqrt 2\, \delta t\sin k ,m\Bigl)\cdot
\v S \right]\Psi_k\nn
&\equiv\sum_k \Psi_k^\dag {\cal H}^\mathrm{I}_k \Psi_k,
\end{align}
where $\Psi_k^T = (\psi^{a}_k,\psi^{c}_k,\psi^{b}_k)$. Definition of the Fourier modes are
$a_{2n}=\frac{1}{\sqrt N} \sum_k e^{-2i k n } a_k$ (similarly $b_{2n+1}$ and $c_{2n}$),
where $N$ is number of $a$-sites (equivalently, $b$ or $c$-cites).
Representation of the spin matrix $\v S$ is such that $S^z$ is diagonal, with entries $+1, 0, -1$.

A second class of topological three-band Hamiltonians is found by considering flux models.
Introducing the diagonal hopping between $a$ and $c$ sites creates triangles that may be threaded
with internal flux, despite the overall lattice being one-dimensional. Choosing the real-space
hopping patterns as shown in Fig. \ref{fig:1}(c) results in the Hamiltonian

\begin{align}\label{h1k2}
H^\mathrm{II}&=\sum_k \Psi_k^\dag \left[2 \Bigl( J_1\cos\left(k\!+\!\lambda\right),
J_2\cos k , J_3 \Bigr)\cdot \v S\right] \Psi_k\nn
&\equiv\sum_k \Psi_k^\dag {\cal H}^{\rm II}_k \Psi_k .
\end{align}
This time, $\Psi_k^T = (\psi^{a}_k,\psi^{b}_k, \psi^{c}_k)$ and the spin matrices are given by $(S^\alpha )_{\beta\gamma} = -i
\varepsilon_{\alpha\beta\gamma}$. Flux penetrating the two triangles
are $\pm \pi/2 +\lambda$, respectively~\cite{comment2}.

Both constructions lead to the structure ${\cal H}_k=\v d_k\cdot \v S$, in different
representations of the spin matrices. Mathematically they may be related by a unitary
transformation, but physically $H^{\rm I}$ and $H^{\rm II}$ represent quite different
situations (modulated real-valued hopping vs. complex-valued hopping, on-site energy
difference vs. diagonal hopping, etc.). It is encouraging that very different physical
conditions can result in the same general Hamiltonian. In both models the first Brillouin
zone extends over $-\pi/2< k \leq\pi/2$, although the periodicity of the Hamiltonian is
$2\pi$: ${\cal H}^{\rm I, II}_{k+2\pi} = {\cal H}^{\rm I, II}_k$. A related observation
was made for two-dimensional topological three-band models earlier~\cite{varma,GPH}, and
readers may consult them regarding this point.

Topological quantum numbers associated with the band Hamiltonians $H^{\rm I}$ and $H^{\rm
II}$ can be readily computed in terms of a two-component unit vector $\hat{n}^a (k) = \v
n^a  (k)/|\v n^a (k)| \equiv (\cos \theta^a_k , \sin \theta^a_k )$, as the integral $N^a
= (1/2\pi) \int_{-\pi}^\pi (d\theta^a_k /dk)$~\cite{1D-kitaev}:

\ba \v n^\mathrm{I} (k) =\left(t\cos k , \delta t \sin k \right), &\!&\! N^{\rm I} =\mathrm{sgn}(t \delta t),  \nn
\v n^\mathrm{II} (k) =\left(J_1\cos (k\!+\!\lambda), J_2\cos k \right),&\!&\! N^{\rm II} =
\mathrm{sgn}(\lambda J_1 J_2 ) . \nn \ea

\begin{figure}[ht]
\includegraphics[width=75mm]{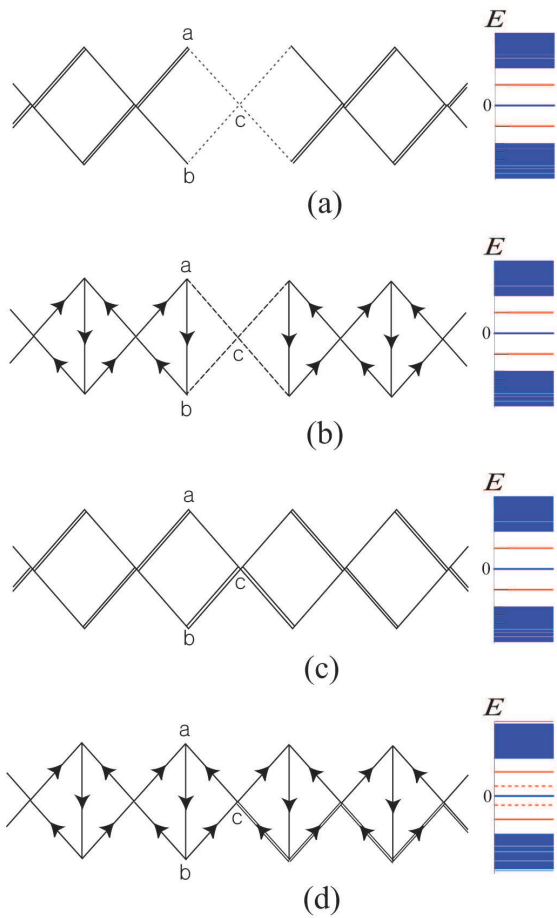}
\caption{(color online) Imposing open boundary condition on the models (a) $H^{\rm I}$
and (b) $H^{\rm II}$. Hopping amplitudes along the dotted lines are zero. A pair of
soliton states are induced on either side of the boundary at energies indicated by red
lines in the schematic energy diagram on the right. The center blue line is the flat band
at zero energy. (c) Reversing the sign of $\delta t$ in $H^{\rm I}$ on half of the
lattice sites in a closed chain. One soliton state per sign-reversed site is induced at
red-lined energies. (d) Reversing the sign of $J_1$ on half the lattice sites in $H^{\rm
II}$. Red-dashed energies are split off from the central flat band by perturbative
effects and do not represent soliton states of our interest.}\label{fig:soliton}
\end{figure}

\textit{Boundary states}.- In a physical system with nontrivial topological index there may exist
localized states near boundaries. In both models with open boundaries as shown in Fig.
\ref{fig:soliton}(a) and (b), there are pairs of localized defect states, one per boundary, at
energies

\begin{align}\label{eq:rho}
&E_s^{\rm I} = \pm 2m,\quad E_s^{\rm II}= \pm 2 J_3,
\end{align}
with the localization factors ($\psi^{a,b,c}_{2j}=\rho^j\psi^{a,b,c}_0$)

\begin{align}\label{eq:rho}
\rho^{\rm I}=-\frac{t^2-\delta t^2}{(t\pm\delta t)^2},\quad
\rho^{\rm II}=-\frac{J_1^2 e^{2i\lambda}+J_2^2}{J_1^2+J_2^2\mp 2
J_1J_2\sin\lambda}.
\end{align}
The $\pm$ sign in Eq. (\ref{eq:rho}) gives either the left- or the right-localized mode
depending on signs of $t\,\delta t$, $J_1 J_2$ and $\lambda$~\cite{SI}. Having found some
localized states, the next obvious questions are whether these localized states represent
the generalization of the familiar Su-Schrieffer-Heeger soliton state~\cite{SSH} to
the three-band model, and, if so, what is the associated charge of the soliton? These are
issues that cannot be answered within the tight-binding analysis alone. For this we turn
to the continuum theory as pioneered by the field theorists~\cite{Jackiw:1975fn,
Jackiw:1982sm,Goldstone:1981kk} and show how they can be generalized to solitons arising
in the three-band model.
\\

\textit{Continuum theory of the three-band soliton}.- In contrast to the two-band
Su-Schrieffer-Heeger soliton state that has received enormous scrutiny in terms of
continuum theory, there appears little, if at all, attempt in the literature to construct
a field-theoretical continuum description of the three-band defect such as found here. We
aim at establishing such continuum theory now and use it to compute the charge of the
soliton. Only one of the Hamiltonians $H^{\rm I}$ and $H^{\rm II}$ needs to be analyzed
for this purpose as the calculated soliton charge will be invariant under the unitary
transformation. We choose $H^{\rm I}$, re-scale $t \rightarrow t/\sqrt2$ and  $\delta t
\rightarrow \delta t /\sqrt 2$, and expand it around $k=\pi/2$ where one obtains both the
maximum topological density $|d\theta^{\rm I}_k /dk|$ and the minimum gap. Choosing
$t=1$, one finds the real-space Hamiltonian

\begin{align}\label{ham2}
{\cal H}= 2\Bigl[ (i\partial_x ) S^x + \delta t S^y + m S^z \Bigr],
\end{align}
where the superscript I has been dropped. Particle-hole symmetry is generally absent in
${\cal H}$ with both $\delta t$ and $m$ finite~\cite{RM}. Soliton state occurs thus at
non-zero energies, either above the flat band or below, depending on the sign of the
kink.

In the case where $\delta t (x)$ behaves as a solitonic background,
$\delta t (x) = \delta t_0 \,{\rm sgn} (x)$, $m(x) = m_0$, (un-normalized) continuum wave
function  of the localized state is found~\cite{SI}

\begin{align}\label{lsol}
&\psi(x) \sim \left(
          \begin{array}{c}
             e^{- \delta t_0 |x|} \\
             0 \\
             0 \\
          \end{array}
        \right),\qquad E_s= \, 2m_0 \quad(\delta t_0>0),
        \nn
&\psi(x) \sim \left(
          \begin{array}{c}
             0 \\
             0 \\
             e^{ \delta t_0 |x|} \\
          \end{array}
        \right),\qquad E_s= \, -2m_0 \quad(\delta t_0<0).
\end{align}
The above solution, derived for the sign-changing $\delta t(x)$ corresponding to Fig. \ref{fig:soliton}(c),
remains also a solution for the case in Fig. \ref{fig:soliton}(a) because identical boundary conditions
($\psi^b_{2j} = 0$) are met by the lattice soliton solutions in both (a) and (c).

\begin{center}
\begin{figure}[ht]
\includegraphics[width=80mm]{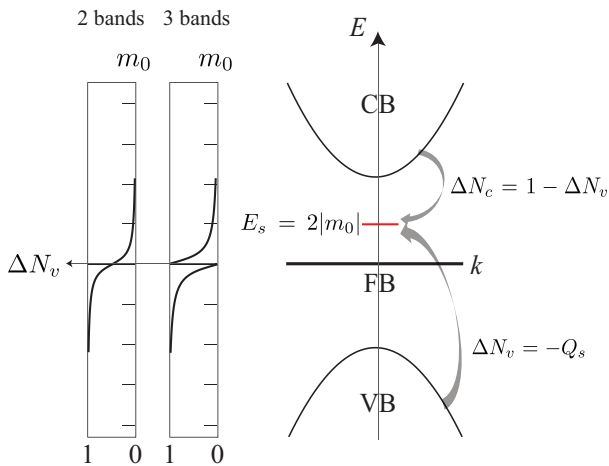}
\caption{(color online) Fractional state lost from the valence band $\Delta N_v$ for two-
and three-band models as a function of the soliton energy on the vertical axis. A
schematic picture on the right shows how the conduction band (CB) and the valence band
(VB) contribute fractional states $\Delta N_c$ and $\Delta N_v$ adding up to one soliton
state, $\Delta N_c + \Delta N_v =1$. Flat band (FB) does not lose its states to the
soliton. } \label{fig:charge}
\end{figure}
\end{center}

We were able to adapt the method of Ref.~\cite{Jackiw:1982sm} to
calculate the soliton charge of the state found in Eq.
(\ref{lsol})~\cite{SI}. The amount of fractional state lost from the
valence band to the soliton, $\Delta N_v$, equals

\begin{align}\label{lsol}
&\Delta N_v =\frac{2}{\pi}\tan^{-1}\left( \left| { \delta t_0  \over m_0 }
\right|\right),~E_s= \, 2|m_0|,~(m_0 \delta t_0>0),\nn &\Delta N_v
=\frac{2}{\pi}\tan^{-1}\left( \left| {m_0 \over \delta t_0} \right| \right),~E_s= \,
-2|m_0|,~(m_0 \delta t_0<0).
\end{align}
The loss from the conduction band $\Delta N_c$ is one minus this number, and together
contributes $\Delta N_v + \Delta N_c = 1$ toward the formation of one soliton defect
while the flat band does not give up any state to it. Soliton charge $Q_s$ is minus the
fractional loss from the valence band: $Q_s = -\Delta N_v$. A corresponding two-band
model, obtained by replacing $S=1$ spin operator $\v S$ by the Pauli matrix $\bm \sigma$
in Eq. (\ref{ham2}), yields $\Delta N_v = 1/2- (1/\pi) \tan^{-1}( m_0 /\delta t_0
)$~\cite{Jackiw:1982sm}. The two calculated soliton charges, for two- and three-band
topological models, are summarized pictorially in Fig. \ref{fig:charge}.

The amount of soliton charge abruptly changes from 0 to 1 at $m_0=0$ even though the
soliton energy $E_s$ changes continuously through zero. This stands in sharp contrast to
the two-band soliton whose fractional charge smoothly varies through 1/2 at the
particle-hole symmetric point $m_0 =0$. Such discontinuous change of the soliton charge
is a distinct feature of the three-band topological model, and sets it apart from the
well-known two-band case. The presence of the central flat band, as required by the
three-band rather than the two-band character of the model, plays a critical role in the
determination of the soliton charge although by itself it does not contribute any states
to its formation. For this reason, we believe the reduction of the three-band Hamiltonian
$\v d_{\v k} \cdot \v S$ to its two-band counterpart $\v d_{\v k} \cdot \bm \sigma$, by
somehow integrating out the flat band, is unlikely.

\textit{Quantum field theory calculation of soliton charge}.- Computation of the soliton
charge above was carried out using the knowledge of quantum-mechanical wave functions in
the soliton background. Generalization of the classic Jackiw-Rebbi result to the
three-band soliton is thus made feasible. On the other hand Goldstone and Wilczek
(GW)~\cite{Goldstone:1981kk} showed how to compute the induced soliton charge using the
quantum field theory technique. In the following we show that GW's approach, like the
wave function method, can be generalized to three-band models.

One begins by writing down a one-dimensional Lagrangian

\begin{align}\label{Lo}
L=\int dx\, \psi^\dag (i \partial_t - i S^x \partial_x - \phi_1  S^z - \phi_2 S^y)\psi ,
\end{align}
where, as in GW's formulation, $(\phi_1, \phi_2 )$ undergo a rotation by angle
$\theta(x)$: $( \phi_1 (x), \phi_2 (x) )= M (\cos \theta (x), \sin \theta (x))$. Current
operator for the fermion field $\psi$ can be written down,  $j^\mu = \psi^\dag \Gamma^\mu
\psi$, with a pseudo-$\Gamma$ matrix $\Gamma^0=\mathbb{I}$, $\Gamma^1=S^x$. Calculating
the average current using the standard technique~\cite{SI} yields

\begin{align}\label{current}
\langle j^\mu \rangle  = -\frac{1}{\pi} \epsilon^{\mu\nu} \partial_\nu \left[  \tan^{-1}
\left(\frac{\phi_2}{\phi_1}\right) \right],
\end{align}
which is \textit{twice} the GW result. In the case where $\phi_2(x)$ behaves as a
solitonic background, i.e. $\phi_2(x)\rightarrow \pm \delta t_0$ as $x \rightarrow \pm
\infty$ while $\phi_1(x)=m_0$ stays constant, we obtain

\begin{align}
Q_s = - \frac{2}{\pi} \tan^{-1} \left(\frac{\delta t_0}{m_0}\right).
\end{align}
from integrating $\langle j^0 \rangle$ over $x$. The result agrees with our previous
quatum-mechanical calculation of the soliton charge in Eq. (\ref{lsol}). Basically this
is twice the charge obtained for the two-band Dirac model and as a result the charge
approaches unity (rather than half as in the Dirac model) at the particle-hole symmetric
point $m_0 = 0$. Since it is the fractional part of the soliton charge which matters, the
soliton charge has to drop discontinuously to zero as $m_0$ passes through zero, as shown
in Fig. \ref{fig:charge}.
\\

\textit{Discussion}.- Defects in topological three-band Hamiltonian $H_{k} = \v d_{k}
\cdot \v S$ in one dimension were examined with emphasis on developing a three-band
analogue to the well-known Jackiw-Rebbi-Su-Schrieffer-Heeger soliton physics in
one-dimensional Dirac bands. All the classic results of Jackiw-Rebbi and
Goldstone-Wilczek are generalized to the three-band model.  We suggest that soliton
states carrying irrational quantum numbers may well exist in compounds that embody the
crystal structure of the diamond-chain lattice.  As a possible candidate material we
point out that the $\alpha$-form of Palladium dichloride (PdCl$_2$) possesses the
diamond-chain structure resembling our model~\cite{PdCl2}. Analysis carried out in this
paper are restricted to a special kind in possession of the central flat band.
Re-locating the flat band, say, to be the top band requires going through a quantum phase
transition that will also qualitatively change the character of the model, let alone the
nature of solitons. It is likely the three-band model whose flat band is located between
the valence and the conduction band represents a qualitatively different physical
situation from the one where it is located at the top or the bottom of the band
structure. The latter case may be reducible to the two-band model plus a trivial flat
band, while the former, as we showed through calculation of the soliton charge,
represents a new class of models.

\acknowledgments We acknowledge fruitful discussions with Alex Altland, Leon Balents,
Akira Furusaki, and Hosho Katsura. J. H. H. is supported by the NRF grant (No.
2013R1A2A1A01006430).


\begin{thebibliography}{99}
\bibitem{comment} The most general three-band Hamiltonian contains eight parameters, one for each Gell-Mann matrix.
Analysis of the topological properties in this general case is currently in progress.
\end{thebibliography}
\end{document}


\title{Soliton Defects in One-dimensional Topological Three-band Hamiltonian: Supplemental Information}

\author{Gyungchoon Go}
\affiliation{Department of Physics, Sungkyunkwan University, Suwon
440-746, Korea}
\author{Kyeong Tae Kang}
\affiliation{Department of Physics, Sungkyunkwan University, Suwon
440-746, Korea}
\author{Jung Hoon Han}
\affiliation{Department of Physics, Sungkyunkwan University, Suwon
440-746, Korea} \affiliation{Asia Pacific Center for Theoretical
Physics, POSTECH, Pohang, Gyeongbuk 790-784, Korea}

\maketitle

\section{Lattice calculation of the boundary state}
In this section we discuss how to obtain the localized solutions at the open boundaries of the two
models,

\begin{align}
H^\mathrm{I}&=\sum_k \Psi_k^\dag \left[2\Bigl(\sqrt 2\, t\cos k ,\sqrt 2\, \delta t\sin k
,m\Bigl)\cdot \v S \right]\Psi_k ,
\nn H^\mathrm{II}&=\sum_k \Psi_k^\dag \left[2 \Bigl(
J_1\cos\left(k\!+\!\lambda\right), J_2\cos k , J_3 \Bigr)\cdot \v S\right] \Psi_k.
\end{align}
In real space the first Hamiltonian becomes

\begin{align}\label{h1r1}
H^{\rm I}=&\sum_{n=0}^N \Bigg[ m \Big(a^\dag_{2n}a_{2n}-c^\dag_{2i}c_{2n}\Big) \nn
&\quad+(t\!-\!\delta t)\Bigl( a^\dag_{2n} b_{2n+1} \! +\! b^\dag_{2n+1} c_{2n+2}\! +\!h.c.\Bigl) \nn
&\quad+ (t\! +\!\delta t)\Bigl( b^\dag_{2n+1} c_{2n} \! +\! b^\dag_{2n+1} a_{2n+2}\! +\! h.c.\Bigl)
\Bigg].
\end{align}
Using the one-particle wave function

\begin{align}\label{1pts}
|\Psi\rangle = \sum_{j} \left(\psi^a_{2j} a^\dag_{2j} + \psi^b_{2j+1} b^\dag_{2j+1}
+\psi^c_{2j} c^\dag_{2j}\right)|0\rangle,
\end{align}
and the Schr\"{o}dinger equation $H^{\rm I} |\Psi\rangle = E^{\rm I} |\Psi\rangle$ we obtain the equations

\ba \label{harpeq} E^{\rm I} \psi^a_{2j}&=&m \psi^a_{2j} + (t-\delta t) \psi^b_{2j+1} +(t+\delta t)
\psi^b_{2j-1},\nn E^{\rm I}\psi^b_{2j+1}&=&(t-\delta t) \psi^a_{2j}+(t+\delta t) \psi^a_{2j+2}
\nn
&& +(t-\delta t) \psi^c_{2j+2}+(t+\delta t) \psi^c_{2j},\nn E^{\rm I}\psi^c_{2j}&=&-m \psi^c_{2j} +
(t-\delta t) \psi^b_{2j-1} +(t+\delta t) \psi^b_{2j+1}. \nn \ea
In order to obtain the boundary states we choose the open boundary conditions

\begin{align}\label{obd}
\psi^b_1= \psi^b_{2N-1}=0,
\end{align}
and insert it into \eqref{harpeq}. There are four equations satisfied at the boundaries:

\ba \label{harpeq2} E^{\rm I} \psi^a_{2} &=& m \psi^a_{2} + (t-\delta t) \psi^b_{3},\nn E^{\rm I}
\psi^c_{2}&=&-m \psi^c_{2} +(t+\delta t) \psi^b_{3},\nn E^{\rm I}\psi^a_{2N-2}&=& m \psi^a_{2N-2} +
(t+\delta t) \psi^b_{2N-3},\nn E^{\rm I}\psi^c_{2N-2}&=&-m \psi^c_{2N-2} +(t-\delta t)
\psi^b_{2N-3}. \ea
For the exponentially localized solution we assume the following ansatz

\begin{align}\label{anlosol}
\psi^{a,c}_{2j+2}=\rho^j\psi^{a,b}_2, \qquad \psi^{b}_{2j+3}=\rho^j\psi^{b}_3 .
\end{align}
Using \eqref{harpeq}, \eqref{harpeq2}, and \eqref{anlosol} we obtain

\begin{align}
&E^{\rm I}= 2 m,\quad \rho^{\rm I} =-\frac{t^2-\delta t^2}{(t+ \delta t)^2},\quad \psi^b_{2j}
=\psi^c_{2j+1}=0,\nn &E^{\rm I} = -2 m,\quad \rho^{\rm I}=-\frac{t^2-\delta t^2}{(t- \delta
t)^2},\quad \psi^a_{2j}=\psi^c_{2j+1}=0.
\end{align}
If one solution is left-localized $(|\rho| < 1)$  the other one $(|\rho| > 1)$ must be localized at
the right boundary. By the same manipulation one can easily derive the boundary state of the second
model $H^{\rm II}$

\begin{align}
E^{\rm II}= \pm 2 J_3, \quad \rho^{\rm II}=-\frac{J_1^2 e^{2i\lambda}+J_2^2}{J_1^2+J_2^2\mp 2
J_1J_2\sin\lambda},\quad \psi^c_{2j+1} =0.
\end{align}

\section{Continuum soliton solution}
Here we work out the soliton solution of the continuum Hamiltonian

\begin{align}
{\cal H}^{\rm I} &= 2\left( S^x i\partial_x  +  \delta t(x) S^y + m(x) S^z\right)\nn
&=2 \left(
    \begin{array}{ccc}
      m & \frac{-i}{\sqrt2}(\delta t(x) \!-\! \partial_x) & 0 \\
      \frac{i}{\sqrt2}(\delta t(x) \!+\! \partial_x) & 0 & \frac{-i}{\sqrt2}(\delta t(x) \!-\! \partial_x) \\
      0 & \frac{i}{\sqrt2}(\delta t(x) \!+\! \partial_x) & -m \\
    \end{array}
  \right),
\end{align}
with the background domain-wall profile, $\delta t(x)={\rm sgn}(x)\delta t_0$ and $m(x)= m_0$.
Solving the Schr\"{o}dinger equation, ${\cal H}^{\rm I} \psi(x)=E\psi(x)$, where $\psi^T(x)=(u(x),v(x),w(x))$,
we have the three equations

\ba \label{se1} && \frac{-i }{\sqrt 2}(\delta t(x)-\partial_x) v = (E-m) u \nn
&& \frac{i}{\sqrt2} (\delta t(x)+\partial_x) v = (E+m) w , \nn
&& \frac{i}{\sqrt2}(\delta t(x)+\partial_x) u  -\frac{i}{\sqrt2}(\delta t(x)-\partial_x) w = E v .
\ea
For the exponentially localized solution we take

\begin{align}\label{ela}
u(x) = u_0 e^{-\kappa |x|}, v(x) = v_0 e^{-\kappa |x|},  w(x) = w_0 e^{-\kappa |x|}.
\end{align}
Inserting \eqref{ela} into \eqref{se1} we obtain a pair of solutions,

\begin{align}
E=m, \qquad \delta t_0=\kappa, \qquad w_0 =v_0 =0, \nn {\rm or} \qquad E=-m, \qquad \delta t_0=-\kappa, \qquad u_0 =v_0 =0.
\end{align}

\section{Calculation of the soliton charge}

Calculating the amount of states missing from the scattering states,
which instead contributed to the formation of a soliton bound state,
is a tricky issue. We found it advisible to build up both intuition
and the necessary technology for the charge calculation with a
familiar example of one-dimensional Schr\"{o}dinger equation,
adapted later to the problem of soliton charge in the three-band
case.

Take the Schr\"{o}dinger problem

\ba [-\partial_x^2 + 2\kappa \delta (x) ] u (x) = \varepsilon u(x)
\ea
allowing for either signs of the delta-potential, $\kappa >0$ or
$\kappa < 0$. The scattering states are constructed as

\ba u_L (x) &=& e^{ikx} + R_k e^{-ikx} \nn u_R (x) &=& T_k e^{ikx}
\ea
where L/R refer to the left/right of the delta-potential. Matching
the boundary conditions readily yields

\ba T_k = \frac{ik}{ik-\kappa} ,\qquad R_k =
\frac{\kappa}{ik-\kappa}  \ea
The normalization of the wave function is such that
$\int_{-L/2}^{L/2} |u (x) |^2 dx = L$, instead of 1. We ignore small
corrections of order $O(1/L)$ in the evaluation of the normalization
factor. The scattering states in the presence of the delta-potential
will be denoted $u^{(\delta)}(x)$, and those in its absence,
$u^{(0)}(x)$. Naively, one might think the number of missing
scattering states is obtained by the integral

\ba \Delta Q \equiv \int_{-\infty}^\infty {dk \over 2\pi}
\int_{-L/2}^{L/2} dx ~ \left( | u_k^{(\delta)}(x) |^2 -
|u_k^{(0)}(x) |^2\right) .\ea
Direct insertion of the wave functions obtained earlier with the $x$
integral $\int_0^\infty dx\,e^{ikx} = i/k+ \pi\delta(k)$ leads to
the formula

\ba \Delta Q =-\int_{-\infty}^\infty {dk \over 2\pi}
{\mathrm{Im}[R_k ] \over k} \!-\!{1\over 2}= \frac{{\rm
sgn}(\kappa)}{2}-\frac12. \ea
It is obvious that $\Delta Q = -1$ for the attractive case
$(\kappa<0)$ and $\Delta Q = 0$ for the repulsive one $(\kappa>0)$.
It simply counts whether a bound state has been formed by the
potential or not, and it is only the attractive case that will
generate one exponentially bound state.

Backed by this exercise, we will study how the state counting goes
for the three-band problem. The eigenvalue problem one needs to
solve is

\ba\label{seuvw}
m u_{k, \varepsilon_k} - \frac{i}{\sqrt2}(\delta t-\partial_x) v_{k, \varepsilon_k} &=& \varepsilon_k u_{k, \varepsilon_k} ,\nn
\frac{i}{\sqrt2}(\delta t +\partial_x) u_{k, \varepsilon_k} - \frac{i}{\sqrt2}(\delta t-\partial_x) w_{k, \varepsilon_k}
&=& \varepsilon_k v_{k, \varepsilon_k} \nn
\frac{i}{\sqrt2}(\delta t+\partial_x) v_{k, \varepsilon_k} - m w_{k, \varepsilon_k}  &=& \varepsilon_k w_{k, \varepsilon_k} \nn \ea
at momentum $k$ and energy $E_k = 2\varepsilon_k$. There may be a
bound-state solution in addition to the scattering solutions
characterized by the wave number $k$. Soliton profile is generally
offered by the functional form $\delta t(x)=\delta t_0 \tanh(\xi x)$, with
$\delta t(x)$ varying from $-\delta t_0$ to $+\delta t_0$ between $x=-\infty$ and
$x=+\infty$. Unfortunately we were not able to find exact scattering
solutions for finite $\xi$. Instead we will consider $\xi\rightarrow
\infty$ where $\delta t (x) = \delta t_0 \mathrm{sgn}(x)$, $\delta t^2(x)=\delta t_0^2$, and
$\partial_x \delta t(x) = 2 \delta t_0 \delta(x)$. This limit is sufficiently
weak so that it gives only one bound state. Solving the
Schr\"{o}dinger equation with the soliton background we obtain

\begin{align}\label{wfsol}
\psi_k(x) =\sqrt{\frac{\varepsilon_k^2-m_0^2}{2\varepsilon_k^2}}
\left(
          \begin{array}{c}
             -\frac{i}{\sqrt2(\varepsilon_k-m_0)} (\delta t - \partial_x) v_k \\
             v_k\\
             \frac{i}{\sqrt2(\varepsilon_k+m_0)} (\delta t + \partial_x) v_k  \\
          \end{array}
        \right),
\end{align}
where $\varepsilon_k=-\sqrt{k^2+\delta t_0^2+m_0^2}$ and $v_k$ is the
properly normalized negative-energy solution satisfying

\begin{align}\label{seu}
\left(-{\partial_x}^2 +\frac{2 m_0 \delta t_0}
{\varepsilon_k}\delta(x)\right)v_k=\left(\varepsilon_k^2-m_0^2-\delta t_0^2\right)v_k,
\end{align}
and $\int_{-L/2}^{L/2} |v_k (x) |^2 dx = \int_{-L/2}^{L/2}
\psi^\dag_k (x) \psi_k (x) dx = L$. The normalization chosen above
is correct up to corrections of order $O(1/L)$ which of course
should vanish in the large-$L$ limit.

For each value of $k$, Eq. \eqref{seu} is equivalent to the
single-particle Schr\"{o}dinger equation subject to the
$\delta$-function potential, thus the existence of the bound state
is determined by the sign of the potential. Since at the moment we
are interested in the negative-energy states, $\varepsilon_k$ is
negative and $v_k^s$ would have no bound state for $m_0 \delta t_0<0$.
Instead, in the case where $m_0 \delta t_0>0$ one expect one bound state
in $v_k^s$. Henceforth we will assume $m_0 \delta t_0>0$.

Integrating the charge density difference
\begin{align}\label{rho}
\rho_k^s(x) - \rho_k^0(x) =\psi_k^{s\dag}(x) \psi_k^s(x) -
\psi_k^{0\dag}(x) \psi_k^0(x),
\end{align}
over all space $x$ and momenta $k$ yields the soliton charge (the
superscript $s$ denotes the soliton system and $0$, the soliton-free
system)~\cite{Jackiw:1982sm}.

\begin{widetext}
\ba\label{solch} \Delta Q &=& \int_{-L/2}^{L/2} dx
\int_{-\infty}^\infty {dk \over 2\pi} [ \rho^s_k (x) - \rho^0_k (x)
] = \int dx \int \frac{dk}{2\pi}
\Bigg\{\left(|v_k^s|^2-|v_k^0|^2\right)\nn &&+
\frac{\varepsilon_k^2+m_0^2}{4\varepsilon_k^2(\varepsilon_k^2-m_0^2)}
{\partial_x}^2 (|v^s_k|^2) +
\frac{m_0}{\varepsilon_k(\varepsilon_k^2-m_0^2)} \partial_x
(\delta t(x)|v^s_k|^2) - \frac{m_0 \delta t_0}{2\varepsilon_k^3} |v^s_k|^2
\delta(x)\Bigg\}. \ea

\end{widetext}
In order to compute the charge difference $\Delta Q$, we solve the
Schr\"{o}dinger equation \eqref{seu}. The solution is given as
\begin{align}\label{asu}
v_k^s(x<0) = e^{ikx} + R_k e^{-ikx},\qquad v_k^s(x>0) = T_k e^{ikx},
\end{align}
where
\begin{align}\label{solu}
T_k =\frac{ik\varepsilon_k }{ik\varepsilon_k - \delta t_0  m_0},\qquad R_k
=T_k -1=\frac{\delta t_0  m_0}{ik\varepsilon_k - \delta t_0  m_0}.
\end{align}
The expression of the coefficients remains valid for both signs of
the energy $\varepsilon_k$.
For the valence band, one can use the dispersion $\varepsilon_k =
-\sqrt{k^2+\delta t_0^2+m_0^2}$ to complete the evaluation of the charge

\begin{widetext}

\begin{align}\label{dqv}
\Delta Q_v&=\int dx \int \frac{dk}{2\pi}\Big( |v_k^s|^2-|v_k^0|^2\Big) + \int \frac{dk}{2\pi}\Bigg[\frac{\delta t_0 m_0}{\sqrt{k^2+\delta t_0^2+m_0^2}(k^2+m_0^2)}(|T_k|^2+|R_k|^2+1)-\frac{\delta t_0 m_0}{(k^2+\delta t_0^2+m_0^2)^\frac{3}{2}} |T_k|^2\Bigg]\nonumber\\
&=-\frac{1}{2} - \int \frac{dk}{2\pi} \frac{1}{k} {\rm Im}(R_k) +\frac{2}{\pi} \tan^{-1}\frac{m_0}{\delta t_0}  -\int \frac{dk}{2\pi}
\Bigg[\frac{\delta t_0 m_0 }{\sqrt{k^2+\delta t_0^2+m_0^2}}\frac{k^2}{(k^2+\delta t_0^2)(k^2+m_0^2)}\Bigg]\nonumber\\
&=-\frac{1}{2} + \frac{2}{\pi} \tan^{-1}\frac{m_0}{\delta t_0}  +
\left[\frac{1}{\pi} \tan^{-1}\frac{\delta t_0}{m_0}+ \frac{1}{\pi} \tan^{-1}\frac{m_0}{\delta t_0}\right] \nonumber\\
&=-1+  \frac{2}{\pi} \tan^{-1}\frac{m_0}{\delta t_0}
= - \frac{2}{\pi} \tan^{-1} \left| \frac{\delta t_0}{m_0} \right|.
\end{align}
\end{widetext}
Here we used $|T_k|^2+|R_k|^2=1$ and $\int_0^\infty dx\,e^{ikx} =
i/k+ \pi\delta(k)$.


%
For the conduction band, on the other hand, similar manipulation
with the positive dispersion $\varepsilon_k =
\sqrt{k^2+\delta t_0^2+m_0^2}$ yields

%

\begin{align}\label{dqc}
\Delta Q_c= - \frac{2}{\pi} \tan^{-1} \left| \frac{m_0}{\delta t_0}
\right|.
\end{align}


From Eqs. \eqref{dqv} and \eqref{dqc} we obtain $\Delta Q_v + \Delta
Q_c =-1$. The vacancy of one state from the continuum bands is
responsible for the occurrence of one mid-gap bound state at energy
$E_s = 2 |m_0|$. The argument, and the derivation, remains unchanged
whether one takes $\delta t_0 >0$ and $m_0 >0$, or the opposite, $\delta t_0 <0$
and $m_0 <0$. In both cases the bound state energy is on the
positive side $E_s= 2 |m_0|$. The soliton charge (more precisely
the loss of fractional states) contributed by each band is given by
the same formulas, Eqs. \eqref{dqv} and \eqref{dqc}.

\section{Quantum field theory calculation of soliton charge}

Here we compute the soliton charge of the model by using quantum
field theory. From the continuum Hamiltonian, we easily derive the Lagrangian
\begin{align}\label{Lo}
L=\int dx\, \psi^\dag (x) (i \partial_t - i S^x \partial_x - \phi_1(x) S^z - \phi_2(x) S^y)\psi(x),
\end{align}
here we replaced $\delta t(x)$ and $m(x)$ to $\phi_2(x)$ and $\phi_1(x)$.
Then we take the unitary transformation

\begin{align}
\psi \rightarrow U \psi, \qquad \psi^\dag \rightarrow \psi^\dag U^\dag,
\end{align}
which transforms

\begin{align}
\phi_1 S^z +\phi_2 S^y \longrightarrow \sqrt{{\phi_1}^2+{\phi_2}^2} S^z \equiv M S^z, \qquad (M\geq0).
\end{align}
We assume $M$ is much larger than the gradients of $\phi_{1,2}$ \cite{Goldstone:1981kk}. The
desired rotation is accomplished by writing $(\phi_1 , \phi_2 ) = M (\cos \theta, \sin \theta)$,

\begin{align}
&U= \exp(i \theta S^x), \quad \theta=\tan^{-1}(\phi_2/\phi_1).
\end{align}
The Lagrangian after the unitary rotation reads

\begin{align}\label{Lt}
L\rightarrow L'&=\int dx\, \psi^\dag (x) (i \partial_t + a_t -i S^x \partial_x -  S^x a_x - M S^z)\psi(x)\nn
&=\int dx\, \psi^\dag (x) \left[i \Gamma^\mu (\partial_\mu - i a_\mu) - M S^z\right]\psi(x)\nn
&=\int dx\, \psi^\dag (x) \left[i \Gamma^\mu \partial_\mu - M S^z\right]\psi(x) \nn
& ~~~~ + \int dx\, \psi^\dag (x) \Gamma^\mu a_\mu \psi(x) = L_0 + L_{int}.
\end{align}
Gauge potentials are defined by $a_\mu=i U^\dag \partial_\mu U= S^x \partial_\mu \theta$. In keeping with the $\Gamma$-matrix notation of the Dirac Hamiltonian we employ a similar notation for $\Gamma^0=\mathbb{I}$, $\Gamma^1=S^x$.

It is  our task to compute the expectation value of the urrent $j^\mu = \psi^\dag \Gamma^\mu \psi$

\ba \label{j3} && \langle j^\mu(x)\rangle  = -i {\rm tr} \left(\Gamma^\mu G_0(0)\right) \nn
&& ~~ -i \int d^2 z {\rm tr}\left( \Gamma^\mu G_0(x-z) \Gamma^\nu a_\nu G_0(z-x)\right) , \ea where
the free propagator $G_0$ is

\begin{center}
\begin{figure}[ht]
\includegraphics[width=70mm]{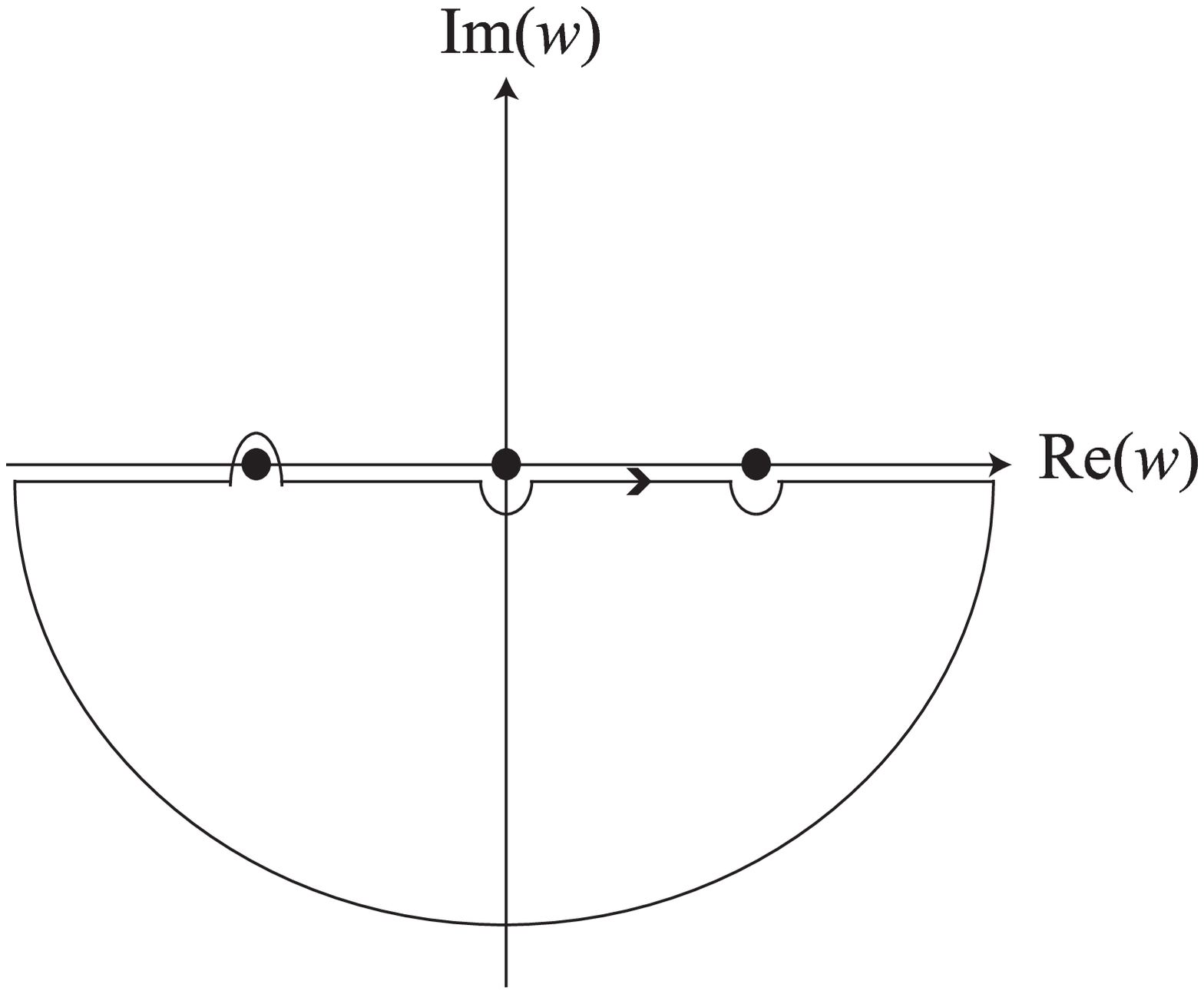}
\caption{Integration contour of $w$.} \label{fig:contour}
\end{figure}
\end{center}

\begin{widetext}
\begin{align}\label{fp}
G_0(x-y) &\equiv {\rm Tr}\left( \frac{i}{-i\partial_t - H_0}\right)\nn
 &= \int \frac{dwdk_x}{(2\pi)^2}\frac{i}{w(w^2-k_x^2-M^2)} \left(\begin{array}{ccc}
    w^2-\frac{k_x^2}{2}+Mw & \frac{k_x w}{\sqrt2} + \frac{k_x M}{\sqrt2}   &  \frac{k_x^2}{2} \\
    \frac{k_x w}{\sqrt2} + \frac{k_x M}{\sqrt2}   & w^2-M^2 & \frac{k_x w}{\sqrt2} - \frac{k_x M}{\sqrt2} \\
    \frac{k_x^2}{2}  & \frac{k_x w}{\sqrt2} - \frac{k_x M}{\sqrt2} & w^2-\frac{k_x^2}{2}- M w \\
  \end{array}\right)e^{i k \cdot (x-y)},
\end{align}
where we use the notation $x_\mu= (t, -x)$, $k_\mu = (w, - k_x)$ and $k\cdot x = k^\mu x_\mu=w t - k_x x$.
The first term of \eqref{j3} reads

\begin{align}
&-i{\rm tr} \left(\Gamma^1 G_0(0)\right) = \int
\frac{dwdk_x}{(2\pi)^2}\left(\frac{2w}{w^2-k_x^2-M^2} \right)=0,\nn &-i{\rm tr} \left(
G_0(0)\right) = \int \frac{dwdk_x}{(2\pi)^2}\left(\frac{1}{w}+\frac{2w}{w^2-k_x^2-M^2} \right)
=\int \frac{dwdk_x}{(2\pi)^2}\left(\frac{1}{w} \right).
\end{align}
It seems that this term gives infra-red (IR) divergence ($w\rightarrow 0$).
The divergence comes from the $E=0$ flat band.
However, since we set the Fermi energy slightly below the flat band ($E_F=0^-$),
the $E=0$ pole is not included in the contour integral Fig.~\ref{fig:contour}.
Thus we have

\begin{align}\label{cd}
\langle j^\mu(q) \rangle=\Pi^{\mu\nu}(q) a_\nu(q) &=-i \int \frac{dwdk_x}{(2\pi)^2} {\rm tr} \left[
\left( {i \over -i\partial_t - H_0 } \right)_{k-q} \Gamma^\mu  \left( {i \over -i\partial_t - H_0 }
\right)_k \Gamma^\nu a_\nu(q) \right].
\end{align}
In the large-$M$ limit, which is
equivalent to the small $q$ limit, the current average becomes

\begin{align}\label{cd2}
\langle j^\mu(q)\rangle \simeq -i \int \frac{dwdk_x}{(2\pi)^2} {\rm tr} \left[ \frac{i}{w - k_x S^x
- M S^z} \Gamma^\mu \frac{i}{w - k_x S^x - M S^z} \Gamma^\nu \Gamma^1 \right] \partial_\nu
\theta(q).
\end{align}
\end{widetext}

\noindent Using \eqref{fp} we compute the traces

\begin{align}\label{tr4}
&{\rm tr} [G_0 \Gamma^1 G_0 \Gamma^1\Gamma^1] = \frac{2k_x (M^2-2w^2)}{w(k_x^2+M^2-w^2)^2},\nonumber\\
&{\rm tr} [G_0 \Gamma^1 G_0 \Gamma^0\Gamma^1] = - \frac{2(k_x^2-M^2+w^2)}{(k_x^2+M^2-w^2)^2},\nonumber\\
&{\rm tr} [G_0 \Gamma^0 G_0 \Gamma^1 \Gamma^1] = - \frac{M^2(k_x^2+M^2-w^2)+2w^2(k_x^2+w^2)}{w^2(k_x^2+M^2-w^2)^2},\nonumber\\
&{\rm tr} [G_0 \Gamma^0 G_0 \Gamma^1 ] = -\frac{4k_x\,w}{(k_x^2+M^2-w^2)^2}.
\end{align}
The first and fourth terms in \eqref{tr4} are odd functions in $k_x$ thus their contributions
vanish after the $k_x$-integration.
Integrating \eqref{cd2} over $w$ and $k_x$ we have

\begin{align}\label{j0}
\langle j^0 \rangle &=-i \int \frac{dk_xdw}{(2\pi)^2} {\rm tr} [G_0 \Gamma^0 G_0 \Gamma^1 \Gamma^1] \partial_1 \theta\nonumber\\
&=i\int \frac{dk_xdw}{(2\pi)^2}\frac{M^2(k_x^2+M^2-w^2)+2w^2(k_x^2+w^2)}{w^2(k_x^2+M^2-w^2)^2}\partial_1 \theta
\nn
& =-\frac{1}{2\pi} \int dk_x \frac{M^2}{(k_x^2+M^2)^{3/2}}\partial_1 \theta \nn
& =-\frac{1}{\pi}\partial_1 \theta,
\end{align}
and

\begin{align}\label{j1}
\langle j^1 \rangle &= -i\int \frac{dk_xdw}{(2\pi)^2} {\rm tr} [G_0 \Gamma^1 G_0 \Gamma^0 \Gamma^1] \partial_0 \theta\nonumber\\
&=i\int \frac{dk_xdw}{(2\pi)^2}\frac{2(k_x^2-M^2+w^2)}{(k_x^2+M^2-w^2)^2}\partial_0 \theta \nn
&=\frac{1}{2\pi} \int dk_x \frac{M^2}{(k_x^2+M^2)^{3/2}}\partial_0 \theta\nonumber\\
&=\frac{1}{\pi}\partial_0 \theta,
\end{align}
where the integration contour for the $w$-integration is taken over the closed path not enclosing
the positive and zero frequency poles as shown in Fig.~\ref{fig:contour}.

Combining \eqref{j0} and \eqref{j1} we obtain

\begin{align}\label{current}
\langle j^\mu \rangle = -\frac{1}{\pi} \epsilon^{\mu\nu} \partial_\nu \theta
= -\frac{1}{\pi} \epsilon^{\mu\nu} \partial_\nu  \tan^{-1} \left(\frac{\phi_1}{\phi_2}\right).
\end{align}
Integrating the $j^0$ over $x$ we easily compute the charge

\begin{align}
Q = - \frac{1}{\pi} \Delta\left( \tan^{-1} \left(\frac{\phi_1}{\phi_2}\right)\right).
\end{align}
In the case where $\phi_2(x)$ behaves as a solitonic background, i.e.
$\phi_2(x)\rightarrow \pm \delta t_0$ as $x \rightarrow \pm \infty$ and $\phi_1(x)=m_0$, we have
\begin{align}
Q = - \frac{2}{\pi} \tan^{-1} \left(\frac{\delta t_0}{m_0}\right).
\end{align}
This result supports our previous quantum mechanical charge computation.